\catcode`\@=11					



\font\fiverm=cmr5				
\font\fivemi=cmmi5				
\font\fivesy=cmsy5				
\font\fivebf=cmbx5				

\skewchar\fivemi='177
\skewchar\fivesy='60


\font\sixrm=cmr6				
\font\sixi=cmmi6				
\font\sixsy=cmsy6				
\font\sixbf=cmbx6				

\skewchar\sixi='177
\skewchar\sixsy='60


\font\sevenrm=cmr7				
\font\seveni=cmmi7				
\font\sevensy=cmsy7				
\font\sevenit=cmti7				
\font\sevenbf=cmbx7				

\skewchar\seveni='177
\skewchar\sevensy='60


\font\eightrm=cmr8				
\font\eighti=cmmi8				
\font\eightsy=cmsy8				
\font\eightit=cmti8				
\font\eightbf=cmbx8				

\skewchar\eighti='177
\skewchar\eightsy='60


\font\ninei=cmmi9
\font\ninesy=cmsy9

\skewchar\ninei='177
\skewchar\ninesy='60


\font\tenrm=cmr10				
\font\teni=cmmi10				
\font\tensy=cmsy10				
\font\tenex=cmex10				
\font\tenit=cmti10				
\font\tensl=cmsl10				
\font\tenbf=cmbx10				
\font\tentt=cmtt10				
\font\tenss=cmss10				
\font\tensc=cmcsc10				
\font\tenbi=cmmib10				

\skewchar\teni='177
\skewchar\tenbi='177
\skewchar\tensy='60

\def\tenpoint{\ifmmode\err@badsizechange\else
	\textfont0=\tenrm \scriptfont0=\sevenrm \scriptscriptfont0=\fiverm
	\textfont1=\teni  \scriptfont1=\seveni  \scriptscriptfont1=\fivemi
	\textfont2=\tensy \scriptfont2=\sevensy \scriptscriptfont2=\fivesy
	\textfont3=\tenex \scriptfont3=\tenex   \scriptscriptfont3=\tenex
	\textfont4=\tenit \scriptfont4=\sevenit \scriptscriptfont4=\sevenit
	\textfont5=\tensl
	\textfont6=\tenbf \scriptfont6=\sevenbf \scriptscriptfont6=\fivebf
	\textfont7=\tentt
	\textfont8=\tenbi \scriptfont8=\seveni  \scriptscriptfont8=\fivemi
	\def\rm{\tenrm\fam=0 }%
	\def\it{\tenit\fam=4 }%
	\def\sl{\tensl\fam=5 }%
	\def\bf{\tenbf\fam=6 }%
	\def\tt{\tentt\fam=7 }%
	\def\ss{\tenss}%
	\def\sc{\tensc}%
	\def\bmit{\fam=8 }%
	\rm\setparameters\setbaselines\fi}


\font\twelverm=cmr12				
\font\twelvei=cmmi12				
\font\twelvesy=cmsy10	scaled\magstep1		
\font\twelveex=cmex10	scaled\magstep1		
\font\twelveit=cmti12				
\font\twelvesl=cmsl12				
\font\twelvebf=cmbx12				
\font\twelvett=cmtt12				
\font\twelvess=cmss12				
\font\twelvesc=cmcsc10	scaled\magstep1		
\font\twelvebi=cmmib10	scaled\magstep1		

\skewchar\twelvei='177
\skewchar\twelvebi='177
\skewchar\twelvesy='60

\def\twelvepoint{\ifmmode\err@badsizechange\else
	\textfont0=\twelverm \scriptfont0=\eightrm \scriptscriptfont0=\sixrm
	\textfont1=\twelvei  \scriptfont1=\eighti  \scriptscriptfont1=\sixi
	\textfont2=\twelvesy \scriptfont2=\eightsy \scriptscriptfont2=\sixsy
	\textfont3=\twelveex \scriptfont3=\tenex   \scriptscriptfont3=\tenex
	\textfont4=\twelveit \scriptfont4=\eightit \scriptscriptfont4=\sevenit
	\textfont5=\twelvesl
	\textfont6=\twelvebf \scriptfont6=\eightbf \scriptscriptfont6=\sixbf
	\textfont7=\twelvett
	\textfont8=\twelvebi \scriptfont8=\eighti  \scriptscriptfont8=\sixi
	\def\rm{\twelverm\fam=0 }%
	\def\it{\twelveit\fam=4 }%
	\def\sl{\twelvesl\fam=5 }%
	\def\bf{\twelvebf\fam=6 }%
	\def\tt{\twelvett\fam=7 }%
	\def\ss{\twelvess}%
	\def\sc{\twelvesc}%
	\def\bmit{\fam=8 }%
	\rm\setparameters\setbaselines\fi}


\font\fourteenrm=cmr12	scaled\magstep1		
\font\fourteeni=cmmi12	scaled\magstep1		
\font\fourteensy=cmsy10	scaled\magstep2		
\font\fourteenex=cmex10	scaled\magstep2		
\font\fourteenit=cmti12	scaled\magstep1		
\font\fourteensl=cmsl12	scaled\magstep1		
\font\fourteenbf=cmbx12	scaled\magstep1		
\font\fourteentt=cmtt12	scaled\magstep1		
\font\fourteenss=cmss12	scaled\magstep1		
\font\fourteensc=cmcsc10 scaled\magstep2	
\font\fourteenbi=cmmib10 scaled\magstep2	

\skewchar\fourteeni='177
\skewchar\fourteenbi='177
\skewchar\fourteensy='60

\def\fourteenpoint{\ifmmode\err@badsizechange\else
	\textfont0=\fourteenrm \scriptfont0=\tenrm \scriptscriptfont0=\sevenrm
	\textfont1=\fourteeni  \scriptfont1=\teni  \scriptscriptfont1=\seveni
	\textfont2=\fourteensy \scriptfont2=\tensy \scriptscriptfont2=\sevensy
	\textfont3=\fourteenex \scriptfont3=\tenex \scriptscriptfont3=\tenex
	\textfont4=\fourteenit \scriptfont4=\tenit \scriptscriptfont4=\sevenit
	\textfont5=\fourteensl
	\textfont6=\fourteenbf \scriptfont6=\tenbf \scriptscriptfont6=\sevenbf
	\textfont7=\fourteentt
	\textfont8=\fourteenbi \scriptfont8=\tenbi \scriptscriptfont8=\seveni
	\def\rm{\fourteenrm\fam=0 }%
	\def\it{\fourteenit\fam=4 }%
	\def\sl{\fourteensl\fam=5 }%
	\def\bf{\fourteenbf\fam=6 }%
	\def\tt{\fourteentt\fam=7}%
	\def\ss{\fourteenss}%
	\def\sc{\fourteensc}%
	\def\bmit{\fam=8 }%
	\rm\setparameters\setbaselines\fi}




\newdimen\rp@
\newcount\@basestretchnum
\newskip\@baseskip
\newskip\headskip
\newskip\footskip


\def\setparameters{\rp@=.1em
	\headskip=24\rp@
	\footskip=\headskip
	\delimitershortfall=5\rp@
	\nulldelimiterspace=1.2\rp@
	\scriptspace=0.5\rp@
	\abovedisplayskip=10\rp@ plus3\rp@ minus5\rp@
	\belowdisplayskip=10\rp@ plus3\rp@ minus5\rp@
	\abovedisplayshortskip=5\rp@ plus2\rp@ minus4\rp@
	\belowdisplayshortskip=10\rp@ plus3\rp@ minus5\rp@
	\normallineskip=\rp@
	\lineskip=\normallineskip
	\normallineskiplimit=0pt
	\lineskiplimit=\normallineskiplimit
	\jot=3\rp@
	\setbox0=\hbox{\the\textfont3 B}\p@renwd=\wd0
	\skip\footins=12\rp@ plus3\rp@ minus3\rp@
	\skip\topins=0pt plus0pt minus0pt}


\def\setbaselines{\maxdepth=4\rp@\baselinestretch=\@basestretchnum}


\def\baselinestretch{\afterassignment\@basestretch\@basestretchnum}
\def\@basestretch{%
	\@baseskip=12\rp@ \divide\@baseskip by1000
	\normalbaselineskip=\@basestretchnum\@baseskip
	\baselineskip=\normalbaselineskip
	\bigskipamount=\the\baselineskip
		plus.25\baselineskip minus.25\baselineskip
	\medskipamount=.5\baselineskip
		plus.125\baselineskip minus.125\baselineskip
	\smallskipamount=.25\baselineskip
		plus.0625\baselineskip minus.0625\baselineskip
	\setbox\strutbox=\hbox{\vrule height.708\baselineskip
		depth.292\baselineskip width0pt }}



\def\makeheadline{\vbox to0pt{\baselinestretch=1000
	\vskip-\headskip \vskip1.5pt
	\line{\vbox to\ht\strutbox{}\the\headline}\vss}\nointerlineskip}

\def\makefootline{\baselineskip=\footskip\line{\the\footline}}

\def\big#1{{\hbox{$\left#1\vbox to8.5\rp@ {}\right.\n@space$}}}
\def\Big#1{{\hbox{$\left#1\vbox to11.5\rp@ {}\right.\n@space$}}}
\def\bigg#1{{\hbox{$\left#1\vbox to14.5\rp@ {}\right.\n@space$}}}
\def\Bigg#1{{\hbox{$\left#1\vbox to17.5\rp@ {}\right.\n@space$}}}


\mathchardef\alpha="710B
\mathchardef\beta="710C
\mathchardef\gamma="710D
\mathchardef\delta="710E
\mathchardef\epsilon="710F
\mathchardef\zeta="7110
\mathchardef\eta="7111
\mathchardef\theta="7112
\mathchardef\iota="7113
\mathchardef\kappa="7114
\mathchardef\lambda="7115
\mathchardef\mu="7116
\mathchardef\nu="7117
\mathchardef\xi="7118
\mathchardef\pi="7119
\mathchardef\rho="711A
\mathchardef\sigma="711B
\mathchardef\tau="711C
\mathchardef\upsilon="711D
\mathchardef\phi="711E
\mathchardef\chi="711F
\mathchardef\psi="7120
\mathchardef\omega="7121
\mathchardef\varepsilon="7122
\mathchardef\vartheta="7123
\mathchardef\varpi="7124
\mathchardef\varrho="7125
\mathchardef\varsigma="7126
\mathchardef\varphi="7127
\mathchardef\imath="717B
\mathchardef\jmath="717C
\mathchardef\ell="7160
\mathchardef\wp="717D
\mathchardef\partial="7140
\mathchardef\flat="715B
\mathchardef\natural="715C
\mathchardef\sharp="715D


\def\err@badsizechange{%
	\immediate\write16{--> Size change not allowed in math mode, ignored}}

\baselinestretch=1000
\tenpoint

\catcode`\@=12					
\catcode`\@=11
\expandafter\ifx\csname @iasmacros\endcsname\relax
	\global\let\@iasmacros=\par
\else	\immediate\write16{}
	\immediate\write16{Warning:}
	\immediate\write16{You have tried to input iasmacros more than once.}
	\immediate\write16{}
	\endinput
\fi
\catcode`\@=12



\def\singlespace{\baselineskip=\normalbaselineskip}
\def\halfspace{\baselineskip=1.5\normalbaselineskip}
\def\doublespace{\baselineskip=2\normalbaselineskip}


\def\AB{\bigskip\parindent=40pt
        \centerline{\bf ABSTRACT}\medskip\halfspace\narrower}
\def\AE{\bigskip\nonarrower\doublespace}
\def\nonarrower{\advance\leftskip by-\parindent
	\advance\rightskip by-\parindent}


\def\boxit#1{\vbox{\hrule\hbox{\vrule\kern3pt
	\vbox{\kern3pt#1\kern3pt}\kern3pt\vrule}\hrule}}

\def\hence{\leavevmode\hbox{\bf .\raise5.5pt\hbox{.}.} }

\def\dalemb#1#2{{\vbox{\hrule height.#2pt
	\hbox{\vrule width.#2pt height#1pt \kern#1pt \vrule width.#2pt}
	\hrule height.#2pt}}}
\def\gtorder{\mathrel{\raise.3ex\hbox{$>$}\mkern-14mu
             \lower0.6ex\hbox{$\sim$}}}
\def\ltorder{\mathrel{\raise.3ex\hbox{$<$}\mkern-14mu
             \lower0.6ex\hbox{$\sim$}}}

\newdimen\fullhsize
\newbox\leftcolumn
\def\twoup{\hoffset=-.5in \voffset=-.25in
  \hsize=4.75in \fullhsize=10in \vsize=6.9in
  \def\fullline{\hbox to\fullhsize}
  \let\lr=L
  \output={\if L\lr
        \global\setbox\leftcolumn=\columnbox\global\let\lr=R \advancepageno
      \else \doubleformat \global\let\lr=L\fi
    \ifnum\outputpenalty>-20000 \else\dosupereject\fi}
  \def\doubleformat{\shipout\vbox{
    \fullline{\box\leftcolumn\hfil\columnbox}\advancepageno}}
  \def\columnbox{\leftline{\vbox{\makeheadline\pagebody\makefootline}}}
  \tolerance=1000 }

\twelvepoint
\doublespace
{\nopagenumbers{
\rightline{IASSNS-HEP-99/85}
\rightline{~~~September, 1999}
\bigskip\bigskip
\centerline{\bf  Derivation of the Lindblad Generator Structure}
\centerline{\bf by use of the It\^o Stochastic Calculus}
\medskip
\centerline{\it Stephen L. Adler
}
\centerline{\bf Institute for Advanced Study}
\centerline{\bf Princeton, NJ 08540}
\medskip
\bigskip\bigskip
\leftline{\it Send correspondence to:}
\medskip
{\singlespace\leftline{Stephen L. Adler}
\leftline{Institute for Advanced Study}
\leftline{Olden Lane, Princeton, NJ 08540}
\leftline{Phone 609-734-8051; FAX 609-924-8399; email adler@ias.edu}}
\bigskip\bigskip
}}
\vfill\eject
\pageno=2
\AB
We use the It\^o stochastic calculus to give a simple derivation of the 
Lindblad form for the generator of a completely positive density matrix 
evolution, by specialization from the corresponding global form for a 
completely positive map.  As a by-product, we obtain a generalized   
generator for a completely positive stochastic density matrix evolution. 

\AE
\bigskip\bigskip
\vfill\eject
\pageno=3
Completely positive maps, and dynamical semigroups constructed from them, 
play an important role in the theory of quantum dissipative systems [1].  
In the global case, the general form of a completely positive map 
$X \to T(X)$  has the simple and intuitive structure [2] 
$$T(X)=\sum_{n\in N} A_n X A_n^{\dagger}~~~,\eqno(1)$$ 
with the $A_n$ operators indexed by $n$.  
In the infinitesimal case, the most general generator of a completely 
positive density matrix evolution has the so-called ``Lindblad form'' [3] 
$${\cal L}\rho=-i[H,\rho] + \sum_{n \in N}[v_n\rho v_n^{\dagger} 
-{1\over 2} \rho v_n^{\dagger}v_n - {1\over 2 } v_n^{\dagger}v_n \rho]~~~,
\eqno(2)$$
with $H$ self-adjoint.  Although one normally expects a simple and evident 
correspondence between the global and infinitesimal forms of 
a transformation, 
the original proofs  of Eq. (2) in Refs. [3]  follow a less direct route.  
Recently, Peres [4] has made the interesting remark that the connection 
between Eq.~(1) and the $v_n \rho v_n^{\dagger}$ term in Eq.~(2) can be 
heuristically understood by identifying $v_n$ with a rapidly 
fluctuating part of $A_n$, of magnitude $(dt)^{1\over 2}$.  
Much the same physical idea is exploited in the 
book of Parthasarathy [5], to give a derivation of Eq.~(2) using stochastic 
calculus methods.  Our aim in this note is to sharpen Peres' observation   
by using the It\^o stochastic calculus to obtain 
Eq.~(2) directly as an infinitesimal specialization of the global 
transformation  of Eq.~(1).  As a by-product, we will obtain a 
generalized form for the generator of an infinitesimal completely positive 
stochastic density matrix evolution.  

We start by specializing Eq.~(1) to operators $A_n$ of the form 
$$\eqalign{
A_n=&d_n+u_ndt +v_n dW_t^n~~~,\cr  
A_n^{\dagger}=&d_n+u_n^{\dagger} dt +v_n^{\dagger} dW_t^n~~~.\cr  
}\eqno(3)$$
Here the $d_n$ are positive real numbers;   no extra generality is achieved 
by taking the $d_n$'s as complex, since by redefining the $u_n,v_n$ a   
phase in $d_n$ can be transformed 
into an overall \break 
c-number phase factor in $A_n$, which 
does not contribute to Eq.~(1).   The $dW_t^n$ are It\^o stochastic 
differentials [6] which obey the algebra  
$$dW_t^mdW_t^n=c^{mn} dt~~,~~dW_t^mdt=0~~~,\eqno(4)$$ 
with $c$ a real symmetric and positive semidefinite covariance matrix.  
By appropriately 
normalizing the operators $v_n$, the diagonal matrix elements of $c$ can 
always be made equal to unity, so that we have 
$$c^{nn}=1~~,~~{\rm all~~} n~~~.\eqno(5)$$
The definition of Eqs.~(4) and (5) includes as special cases that in 
which the It\^o differentials are all the same, $dW_t^n=dW_t,~ {\rm all}~~n$ 
(for which $c^{mn}=1,~{\rm all}~~m,n$), and that in which the It\^o 
differentials $dW_t^n$ are all independent( for which $c^{mn}=\delta^{mn})$.  

Consider now the completely positive density matrix transformation 
$$\rho  \to \rho+d\rho=T(\rho)~~~.\eqno(6)$$
Substituting Eq.~(3) into Eqs.~(1) and (6) and using Eq.~(4), we get 
$$\eqalign{
\rho+d\rho=&\sum_{n\in N} (d_n+u_n dt +v_n dW_t^n)\rho(d_n+u_n^{\dagger}dt 
+v_n^{\dagger}dW_t^n)~~~\cr
=&\sum_{n\in N} d_n^2 \rho 
+ \sum_{n\in N} d_n (v_n \rho + \rho v_n^{\dagger}) dW_t^n 
+(\rho U^{\dagger} + U\rho +\sum_{n\in N} v_n \rho v_n^{\dagger}) dt~~~,\cr 
}\eqno(7a)$$ 
with
$$U=\sum_{n\in N} d_n u_n~~~.\eqno(7b)$$
Equating the coefficients of $\rho$ on the left and right hand sides of 
Eq.~(7a) gives the condition  
$$\sum_{n\in N} d_n^2=1~~~,\eqno(7c)$$
while for the change in $\rho$ we get
$$d\rho=
 \sum_{n\in N} d_n (v_n \rho + \rho v_n^{\dagger}) dW_t^n 
+(\rho U^{\dagger} + U\rho +\sum_{n\in N} v_n \rho v_n^{\dagger}) dt~~~. 
\eqno(8)$$                   
Let us now take the expectation value of Eq.~(8) with respect to the 
stochastic process; since the operators $u_n,~v_n$ have no  
dependence on the It\^o stochastic differentials, and since 
$E[\rho dW_t^n]=0$ in the It\^o calculus [6], we get simply
$$dE[\rho]=E[d\rho]=
\left( E[\rho] U^{\dagger} + UE[\rho] +\sum_{n\in N} v_n 
E[\rho] v_n^{\dagger} \right)dt
~~~.\eqno(9)$$

Now let us impose the condition that the density matrix must always 
have trace unity, which implies 
that ${\rm Tr} d\rho={\rm Tr} E[d\rho]=0$.  From 
Eq.~(9), we get by cyclic permutation under the trace 
$$0={\rm Tr} E[\rho]\left[ U+U^{\dagger}+\sum_{n\in N}v_n^{\dagger} v_n 
\right]~~~,\eqno(10a)$$
which can hold for general $E[\rho]$ only if the coefficient operator 
is zero, which implies that 
$$U+U^{\dagger}=-\sum_{n\in N}v_n^{\dagger} v_n~~~.\eqno(10b)$$
Thus the condition that the infinitesimal transformation preserve the 
trace of $\rho$ determines the self-adjoint part of $U$, while the 
anti-self-adjoint part can be an arbitrary operator $iH$, with $H$ 
self-adjoint, so that $U$ has the form 
$$U=-iH-{1\over 2} \sum_{n \in N}v_n^{\dagger} v_n   ~~~.\eqno(11)$$ 
Substituting Eq.~(11) into Eq.~(9) then gives for the completely positive, 
trace preserving infinitesimal 
deterministic evolution of $E[\rho]$ the expression 
$${dE[\rho] \over dt} = -i[H,E[\rho]]
+\sum_{n\in N}\left[v_n E[\rho] v_n^{\dagger} -{1\over 2} v_n^{\dagger} v_n 
E[\rho] - {1\over 2} E[\rho] v_n^{\dagger} v_n \right]~~~,\eqno(12)$$
which is the Lindblad form.  

Returning now to the stochastic evolution of Eq.~(8), substituting the 
above results and the condition ${\rm Tr}d\rho=0$, we get the additional 
condition 
$${\rm Tr} \rho \sum_{n\in N}d_n(v_n+v_n^{\dagger}) dW_t^n=0~~~,\eqno(13a)$$
which can hold for general $\rho$ only if the operator coefficient vanishes, 
so that 
$$\sum_{n\in N}d_n(v_n+v_n^{\dagger}) dW_t^n=0 ~~~.\eqno(13b)$$      
Multiplying by $dW_t^m$ we get the equations 
$$\sum_{n\in N}d_n(v_n+v_n^{\dagger}) c^{nm}=0~,~~~{\rm all}~~ m\in N~~~.
\eqno(13c)$$
The condition of Eq.~(13c) can be rewritten by noting that since $c^{nm}$ 
is real symmetric and nonnegative, it is diagonalized by an orthogonal 
matrix $O^{mr}$ to yield nonnegative eigenvalues $c^r$, 
$$\sum_{m \in N} c^{nm} O^{mr}=O^{nr}c^r~~~,\eqno(14a)$$ 
with no sum over $r$ on the right.  Thus Eq.~(13c) is equivalent to 
$$\sum_{n\in N}d_n(v_n+v_n^{\dagger}) O^{nr}c^r=0, ~~{\rm all}~~ r\in N~~~.
\eqno(14b)$$  For those values of $r$ for 
which $c^r>0$, we can factor out $c^r$ to 
give the restriction 
$$\sum_{n\in N}d_n(v_n+v_n^{\dagger}) 
O^{nr}=0~~,~~{\rm all~~} r {\rm~~ with~~}  
c^r>0~~~,
\eqno(14c)$$
while for those $r$ for which $c^r=0$, there is no restriction. 
Defining diagonalized It\^o differentials $dZ_t^r$ by 
$$dZ_t^r=\sum_{n\in N}dW_t^n O^{nr}~~~,\eqno(15a)$$
with the inversion 
$$dW_t^n = \sum_{r \in N}O^{nr}  dZ_t^r~~~,\eqno(15b)$$
we find that 
$$dZ_t^qdZ_t^r=\sum_{n,m \in N}dW_t^ndW_t^m O^{nq}O^{mr}
=\sum_{n,m \in N} c^{nm}O^{nq}O^{mr}dt=\delta^{qr}c^rdt~~~,\eqno(15c)$$
and that the restriction of Eq.~(13b) takes the form
$$\sum_{n,r\in N}d_n(v_n+v_n^{\dagger}) O^{nr}dZ_t^r=0 ~~~.\eqno(15d)$$   
Since Eq.~(15c) implies that for those $r$ for which $c^r=0$ the 
corresponding $dZ_t^r$ is idempotent, and hence vanishes, the 
reduced set of restrictions given in Eq.~(14c) suffices 
to guarantee the vanishing of Eq.~(15d)  and hence the satisfaction 
of the original condition of Eq.~(13b).  
We conclude that the completely positive, trace preserving stochastic 
evolution of $\rho$ corresponding to our construction of Eq.~(12) is 
generated by
$$d\rho=\sum_{n\in N}d_n(v_n\rho+\rho v_n^{\dagger}) dW_t^n
+\left[ -i[H,\rho]+\sum_{n\in N}[v_n\rho v_n^{\dagger}
-{1\over 2}v_n^{\dagger}v_n \rho- {1\over 2}\rho v_n^{\dagger}v_n] \right]dt
~~~,\eqno(16)$$
subject to the restrictions on the positive real numbers $d_n$ and the 
operators $v_n$ of Eqs.~(7c) and (14c).  

The simplest case of Eq.~(16) is that in which the sum over the index 
set $N$ contains only one term, so that Eq.~(7c) implies   
$d_1=1$ and Eq.~(5) completely determines the covariance matrix to be 
$c^{11}=1$.  Dropping the superfluous index $n$, 
we get 
$$d\rho=(v\rho+\rho v^{\dagger}) dW_t 
+\left[ -i[H,\rho]+v \rho v^{\dagger}
-{1\over 2} v^{\dagger}v\rho -{1\over 2}\rho v^{\dagger} v  \right] dt
~~~,\eqno(17)$$
with the restriction of Eq.~(14c) simplifying to 
$$v+v^{\dagger}=0~~~,\eqno(18a)$$
which implies that 
$$v=-iK~~~,\eqno(18b)$$ 
with $K$ self adjoint.  Substituting Eq.~(18b)  back into 
Eq.~(17), we get 
$$\eqalign{
d\rho=&-i[K,\rho] dW_t 
+\left[ -i[H,\rho]+K \rho K
-{1\over 2} K^2 \rho -{1\over 2}\rho K^2  \right] dt\cr
=&-i[K,\rho] dW_t 
+\left[ -i[H,\rho]-{1\over 2} [K,[K, \rho]]  \right] dt~~~,\cr
}\eqno(18c)$$
while from Eqs.~(3), (7b),  and (11) we see that  
$$A=1-iKdW_t+\left(-iH-{1\over 2}K^2\right)dt 
=\exp[-iHdt -iK dW_t]~~~.\eqno(19)$$
Thus, in the case when the index set $N$ contains a single term, 
the transformation of Eq.~(16) reduces to an infinitesimal stochastic 
unitary transformation.  

\bigskip
\centerline{\bf Acknowledgments}
This work was supported in part by the Department of Energy under
Grant \#DE--FG02--90ER40542.
\vfill\eject
\centerline{\bf References}
\bigskip
\noindent
[1]  See, e.g., V. Gorini, A. Frigerio, M. Verri, A. Kossakowski,      
and E. C. G. Sudarshan, Reports on Math. Phys. 13 (1978) 149.\hfill\break 
\bigskip                        
\noindent
[2]  K. Kraus, Ann. Phys. (NY) 64 (1971) 311; E. B. Davies, {\it Quantum 
Theory of Open Systems}, Academic Press, London, 1976, Sec. 9.2.\hfill\break
\bigskip
\noindent
[3]  G. Lindblad, Commun. Math. Phys. 48 (1976) 119; V. Gorini, A. 
Kossakowski, and E. C. G. Sudarshan, J. Math. Phys. 
17 (1976) 821.\hfill\break
\bigskip
\noindent
[4]  A. Peres, quant-ph/9906023, Sec. V.\hfill\break
\bigskip
\noindent
[5]  K. R. Parthasarathy, {\it An Introduction to Quantum Stochastic 
Calculus}, Birkh\"auser Verlag, Basel, 1992, Chapt. III.\hfill\break
\bigskip
\noindent                
[6]  For an excellent exposition of the It\^o calculus, see C. W. Gardiner, 
{\it Handbook of Stochastic Methods}, Springer-Verlag, Berlin, 1990, 
Chapt. 4.\hfill\break  

\bye